# Micromagnetic simulations of absoption spectra in magnetic nanodots, r.f. field perpendicular to the samples' plane.

**Abstract**

Recently developed eigenvalue method is being used to calculate absorption spectra in magnetic square nanodots and nanodisks. Obtained results are being compared with both theoretical and experimental results obtained previously for such structures.

**Introduction**

One of the most powerful ways to study properties of magnetic nanostructures is to study their resonant spectra. Various methods can be used to do so – FerroMagnetic Resonance measurements, Brillouin Light Scattering, spintronic excitations[i] in one or another way produce resonant spectra - response of the system as a function of parameters of the applied excitation; for example the energy absorbed as a function of external r.f. field's frequency. While there is a variety of methods one can use in order to observe what kind of oscillations in magnetic system are responsible for the peaks in the absorption spectra (for example, Magneto-Optical Kerr Effect - MOKE), the resolution of most of such methods is relatively small. As a result, in order to fully understand the experimental results one needs to use numerical calculations together with analytical predictions based on approximate continuum theory. Until recently most of numerical experiments have been performed via the famous OOMF NIST code[ii]. Based on using finite differencies method in order to directly solve Landau-Lifshitz equation, this program allows one to obtain the time-dependent response of the system to any external r.f. field. However, one needs to wait for the system to reach the steady state for which the energy absorbed from the r.f. field equals the energy transferred to the system through damping before one is capable to determine the value of energy absorbed per unit time. Any change in parameters of the external r.f. field requires all of the calculations to be repeated. In addition to this, it's somewhat hard to extract the information about the spatial distribution of oscillation modes exited by the external r.f. field.

As an alternative to this scheme, so called eigenvalue method have been developed in parallel by us[iii] and by other groups[iv,v]. These methods allow one to calculate all the resonant modes and resonant frequencies of the given system. Later, we succeeded in developing the technique that allows one to calculate the response of the system to external fields, spin polarized currents and so on[vi]. This technique also allows one to calculate the absorption spectra for arbitrary external r.f. fields.

In the present paper we will use this method to calculate the absorption spectra of magnetic nanostructures – disks and square slabs in out of plane r.f. field.

**Numerical Method.**

Presented here is a brief review of our formalism (a more detailed discussion is given elsewhere[vi,vii]). We recall the Landau-Lifshitz equation in the presence of a dissipative Gilbert term:

$$\frac{d\mathbf{m}_i}{dt} = -\gamma \mathbf{m}_i \times \mathbf{h}_i^{total} - \frac{\beta\gamma}{M_s} \mathbf{m}_i \times \left(\mathbf{m}_i \times \mathbf{h}_i^{total}\right); \tag{1}$$

here $\mathbf{m}_i$ is the magnetic moment of the $i^{th}$ spin, $\gamma$ is the gyromagnetic ratio, and $\beta$ is a parameter governing the dissipation; henceforth a "spin" or "discrete dipole" can imply a variety of objects, from discrete dipoles per se, to close-packed uniformly magnetized prisms. Despite the firm belief of many that there is a drastic difference between using discrete dipoles and uniformly magnetized prisms, if used properly, both methods produce nearly identical results (resonant frequencies, switching times etc.).

Our strategy is to linearize the problem by writing both the applied fields and magnetic moments as the sum of a zeroth-order static part and a small first-order time-dependent perturbation:

$$\mathbf{m}_i = \mathbf{m}_i^{(0)} + \mathbf{m}_i^{(1)}(t) \tag{2a}$$

$$\mathbf{h}_i = \mathbf{h}_i^{(0)} + \mathbf{h}_i^{(1)}(t). \tag{2b}$$

Assuming a solution of the form

$$\mathbf{m}_i^{(1)}(t) = \hat{\mathbf{x}}_i \, V_{i\alpha}^{(k)} \, e^{-i\omega t} \tag{3}$$

and linearizing Eq. (1) with respect to $\mathbf{m}_i^{(1)}$ and $\mathbf{h}_i^{(1)}$ we obtain the eigenvalue equation:

$$i\omega^{(k)} V_{i\alpha}^{(k)} = \gamma B_{ij\alpha\theta} V_{j\theta}^{(k)}$$

$$B_{ij\alpha\theta} = \varepsilon_{\alpha\beta\gamma}\left[m_{i\beta}^{(0)} A_{ij\gamma\theta}^{(1)} + \delta(\beta-\theta;i-j)h_{i\gamma}^{(0)}\right] + \frac{\beta}{M_s}\varepsilon_{\alpha\beta\gamma}\varepsilon_{\gamma\phi\varphi}m_{i\beta}^{(0)}\left[m_{i\phi}^{(0)} A_{ij\varphi\theta}^{(1)} + \delta(\phi-\theta;i-j)h_{i\varphi}^{(0)}\right]$$

$$\tag{4}$$

here and after, the summation convention is going to be used; roman letters correspond to individual dipoles, greek letters correspond to coordinates. $\omega^{(k)}$ and $V_{i\alpha}^{(k)}$ are, respectively, $k^{th}$ eigenvalue and eigenvector of Eq. (4); $A_{ij\gamma\theta}^{(1)}$ is a demagnetization tensor, which accounts for dipole-dipole and exchange interactions and, if necessary, uniaxial anisotropy.

In the presence of external r.f. field Eq. (4) becomes a homogeneous part of a more general inhomogeneous equation, in which external r.f. field is responsible for a source (inhomogeneous) term. This equation can be solved by various methods[vii] in order to produce the following steady-state solution (here we consider only the case where the applied r.f. field has a sinusoidal time-dependence):

$$m_{i\alpha}^{(1)} = -iV_{i\alpha}^{(k)} \frac{V_{l\beta}^{(k)*}\gamma\varepsilon_{\beta\sigma\chi}m_{l\sigma}^{(0)}h_{l\chi}^{(rf)}}{\omega^{(k)} - \omega} e^{-i\omega t} \qquad (5)$$

The energy absorbed per unit time is given by:

$$\dot{E} = \omega V \gamma \, \text{Re}\left(-\frac{V_{l\beta}^{(k)*}\varepsilon_{\beta\sigma\chi}m_{l\sigma}^{(0)}h_{l\chi}^{(rf)}}{\omega - \omega'^{(k)} + i\beta\omega''^{(k)}} V_{i\alpha}^{(k)} h_{i\alpha}^{(rf)*}\right) \qquad (6)$$

here V is the object's volume, $\omega$ is the frequency of the r.f. field, $h_{i\alpha}^{(rf)}$ is $\alpha$ component of the external r.f. field at the discretization point i.

**Absorption Spectra.**

We have applied the method described here to calculate the absorption spectra and resonant modes in two magnetic nanodots: a disk, D = 175 nm diameter, L = 25 nm thickness and a square slab 150×150×25 nm. In both cases we chose permalloy as the material, which has the following parameters: saturation magnetization $M_s = 795 \, \text{emu/cm}^3$, exchange stiffness $A = 1.3 \cdot 10^{-6}$ erg/cm, and damping coefficient $\beta = 0.01$. For individual cells we chose 2.5×2.5×25 nm blocks, which corresponds to studying only the modes that are uniform in the direction perpendicular to the plane of the disc. Consider the case where no external d.c. magnetic field is present; the equilibrium configuration in this case is then a so-called vortex state, which has been described previously in some detail[viii,ix,x]. It is characterized by a localized structure (the vortex core) in the center of a circular or square slab where the spins mostly point out of the plane of magnetization, whereas the remainder of the spins lie on closed loops around the core. In rings and square slabs with a central hole, the core is absent, with the rest of the sample having a distribution of magnetization similar to that in disks.

In our previous article[vii] we addressed in a short form the case when the applied r.f. field is uniform and lies in the sample's plane. In this work we will consider only the r.f. fields that are perpendicular to the sample's plane. We will consider two spatial dependencies – uniform fields and the fields that linearly depend on one of the in-plane coordinates. In the latter case the dependency is such that the r.f. field is 0 in the middle of the sample, while being positive on one side, and negative on the side of the sample.

Theoretically the modes in a disk can be described as[xi,xii,xiii]:

$$m^{(1)} = e^{im\varphi} F_n(r) \qquad (7)$$

where $\varphi$ is in-plane angle, m and n are integer numbers, and $F_n(r)$ is a combination of Bessel functions, depending only on the distance from the disk's center.

The modes sharing the same absolute value of m are supposed to form doublets. The difference in such modes' frequency is due to the presence of the vortex core, and disappears if the vortex core is removed (for example in rings).

From Eq.(6) we know that in order to be excited, projection of the mode onto global Cartesian coordinate system should have similar symmetry properties to that of the applied r.f. field. It means that when the uniform r.f. field is being used, only the modes with uniform or evenly symmetric projection on corresponding coordinate axis are going to be excited. On the other hand, one has to remember that $m^{(1)}$ describes the magnetization precession in a local coordinate system, formed by the equilibrium direction of magnetization; because of this, sometimes it can be hard to identify the excited modes by looking at their projections onto the global coordinate system; instead one has to use the representation in the local coordinates and carefully analyze the phase dependence. However, in order to present the modes (as for example is done in Figure 2) we will use a projection of the modes onto the global coordinate system for out-of-plane coordinate. The color will represent the sign and magnitude of oscillations. We choose this method of presentation because of its similarity to the ones used by the experimentalists. Concerning the spatial dependence of the excited modes, since in the vortex core the equilibrium magnetization is aligned almost perpendicular to the disk's plane, the modes contained within the vortex core can not be effectively excited by out of plane r.f. fields.

When analyzing the numerical results one also has to remember that all numerical methods introduce some kind of approximation; in our case the use of square discretization lattice leads to the following errors: the difference in frequency among doublets is slightly increased, the modes themselves also appear slightly distorted in favor of four fold symmetries.

On Figure 1 we show the absorption spectrum of a disk subjected to uniform and linear r.f. fields, applied perpendicular to the disk's plane and Figure 2 shows some of the most excited modes. Absorption spectrum clearly resembles the experimental results obtained by Zhu et al[xiv].

Similar to results obtained by Giovannini et al[xv] all modes with even m are standing waves with $A\cos(m\varphi) + B\sin(m\varphi)$ instead of $e^{im\varphi}$ angular dependence; all modes with odd m are running waves with $e^{im\varphi}$ angular dependence; another important difference from Eq.(7) is due to the inter-mode coupling, which is responsible for forming the modes with angular dependence of form $e^{im_1\varphi}\cos(m_2\varphi)$, as an example m=0, n=0 mode "coexists" with m=4, n=0 mode. As discussed above, since r.f. field excites the modes with the symmetry properties similar to its own, uniform r.f. fields excites modes with even m, and linear r.f. field excites the modes with odd m.

The mode depicted by Fig. (2.I) is m=0, n=1 mode, the mode in Fig. (2.II) is m=0, n=2 mode; while modes with higher m numbers are also being excited, the dominant part in the spectra belongs to modes with m=0.

When the external field depends linearly, again the modes with lowest possible m number (since m should be odd in order to be excited by r.f. field this number is 1) dominate the spectra; Figures (2.III-2.II.) respectively show the modes characterized by m=1, n=0 (gyroscopic mode[xvi]), 1, 2, 3 and 4.

There is no general theory for the modes of a vortex configuration in a square slab, therefore our analysis of such case is limited. Figure 3 shows the absorption for the uniform and linear out-of-plane r.f. fields. Figure 4 shows some of the most excited modes. As one can see Figure 4.IV. shows the gyroscopic mode; Figures 4.V. and 4.VI. show the "corner" modes – since the vortex configuration in square slabs can be presented as four domains touching each other at the corners, "corner" modes are in fact the modes localized along the domain walls.

**Conclusions.**

Absorption spectra for magnetic square slabs and disks in the presence of uniform and linear out-of-plane r.f. fields have been obtained and analyzed.

The program that incorporates our code is available for public use at www.rkmag.com.

**Acknowledgments.**

This work was supported by the National Science Foundation under grant ESC-02-24210.

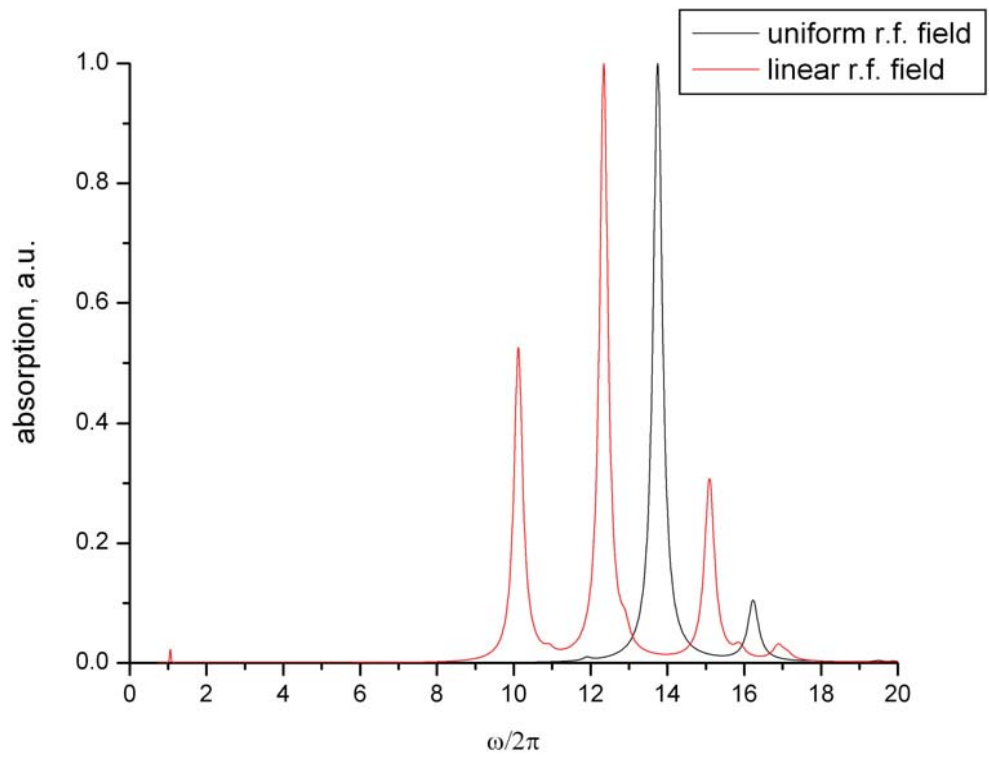

**Figure 1. Absorption spectrum, disk (D = 175nm, L= 25nm); uniform and linear out-of-plane r.f. fields.**

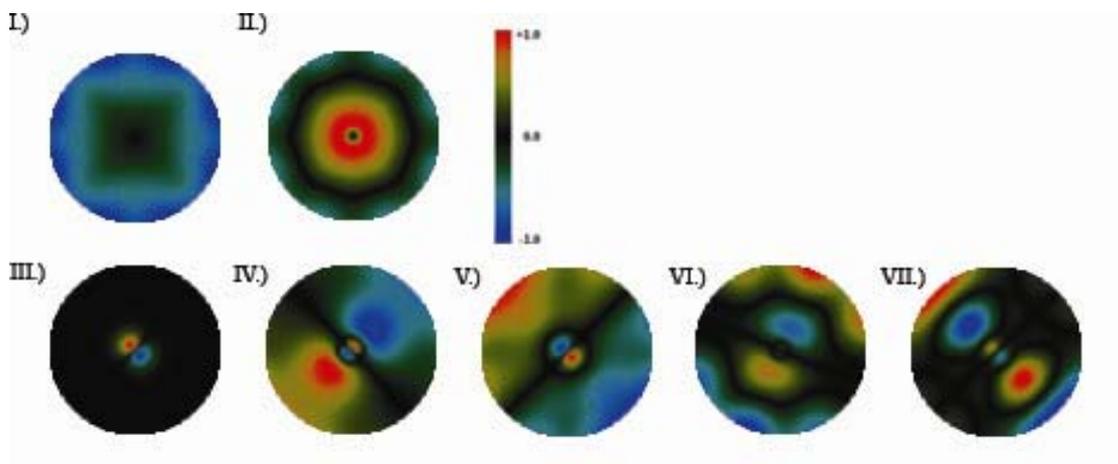

**Figure 2. Some of the most strongly excited modes for a disk (D=175nm, L=25nm); for (I) and (II) r.f. field is uniform, for (III)-(VII) - r.f. field is linear; r.f. field is out-of-plane; the frequencies of the modes are:**

**I.) $\omega/2\pi$=13.74 GHz; II.) $\omega/2\pi$= 16.22 GHz; III.) $\omega/2\pi$=1.05 GHz; IV.) $\omega/2\pi$=10.11 GHz; V.) $\omega/2\pi$=12.34 GHz; VI.) $\omega/2\pi$=15.09 GHz; VII.) $\omega/2\pi$=16.88 GHz.**

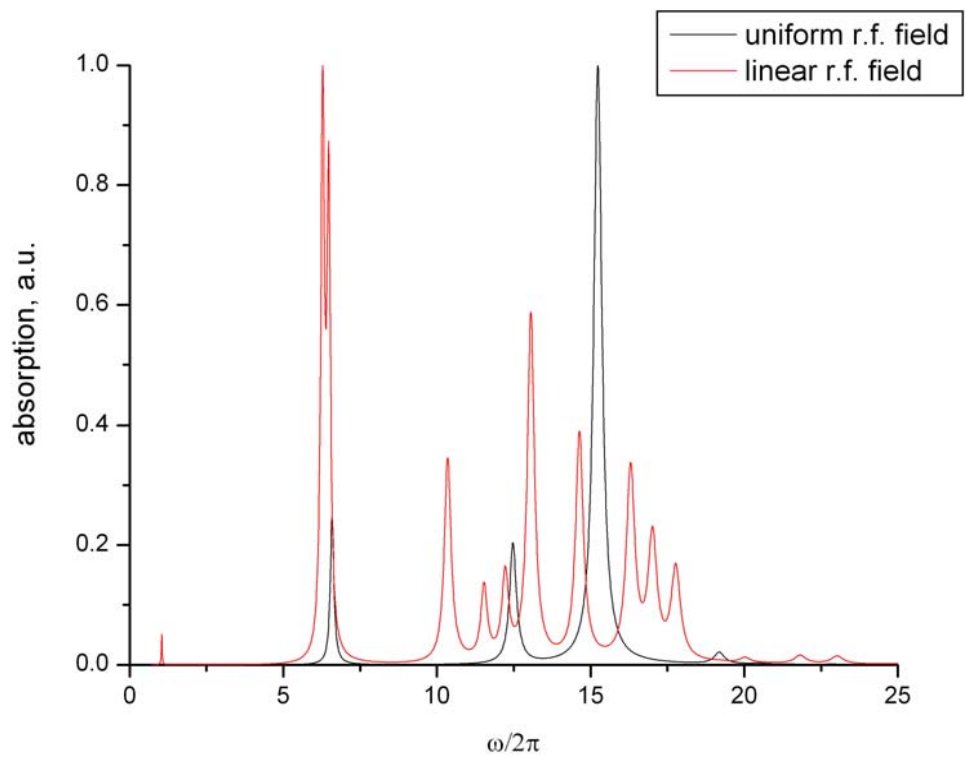

**Figure 3. Absorption spectrum, square slab (150x150x25nm); uniform and linear out-of-plane r.f. fields.**

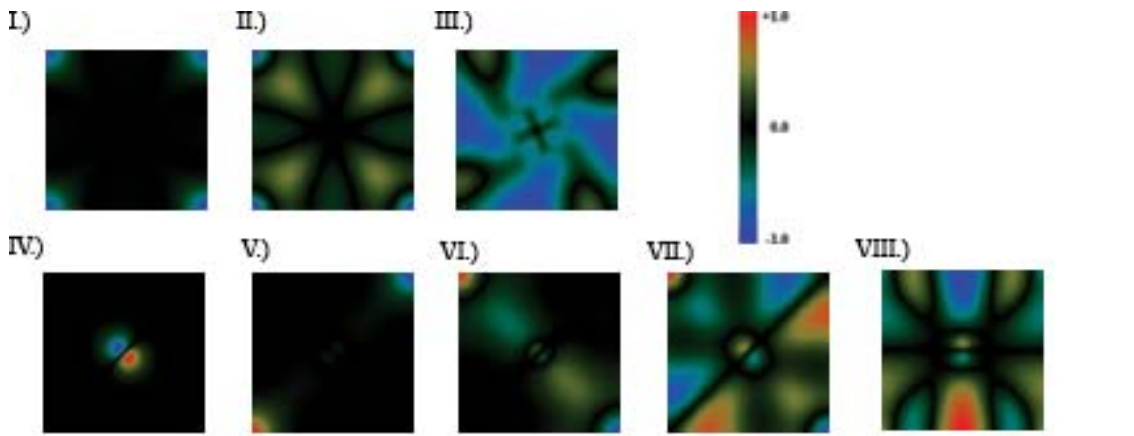

**Figure 4. Some of the most strongly excited modes for a square slab (150x150x25nm); (I) - (III) r.f. field is uniform, for (IV)-(VIII) - r.f. field is linear; r.f. field is out-of-plane; the frequencies of the modes are:**

**I.) $\omega/2\pi$=6.58 GHz; II.) $\omega/2\pi$= 12.47 GHz; III.) $\omega/2\pi$=15.23 GHz; IV.) $\omega/2\pi$=1.04 GHz; V.) $\omega/2\pi$=6.27 GHz; VI.) $\omega/2\pi$=10.34 GHz; VII.) $\omega/2\pi$=13.05 GHz; VIII.) $\omega/2\pi$=14.63 GHz.**